# From Crime to Hypercrime: Evolving Threats and Law Enforcement's New Mandate in the AI Age

Francesco (Frank) Schiliro'
Macquarie University, Sydney Australia
frank.schiliro@mq.edu.au

15 November 2024

**Abstract**

The paper examines the trajectory of crime, tracing its evolution from traditional forms to digital manifestations in cybercrime, and proposes "Hypercrime" as the latest frontier. Leveraging insights from Michael McGuire's "Hypercrime: The New Geometry of Harm," the study calls for a paradigm shift in law enforcement strategies to meet the challenges posed by AI-driven hypercrime. Emphasis is placed on understanding hypercrime's complexity, developing proactive policies, and embracing technological tools to mitigate risks associated with AI misuse.

Portuguese Translation

O artigo examina a trajetória do crime, traçando sua evolução desde as formas tradicionais até as manifestações digitais no crime cibernético, e propõe o "hipercrime" como a última fronteira. Aproveitando os insights de "Hypercrime: The New Geometry of Harm", de Michael McGuire, o estudo pede uma mudança de paradigma nas estratégias de aplicação da lei para enfrentar os desafios colocados pelo hipercrime impulsionado pela IA. A ênfase é colocada na compreensão da complexidade do hipercrime, no desenvolvimento de políticas proativas e na adoção de ferramentas tecnológicas para mitigar os riscos associados ao uso indevido da IA.

## 1. Introduction

Crime, in its various forms, has long mirrored the evolving nature of society. From rudimentary acts of theft and violence in early civilizations to intricate, digitally orchestrated offenses, the trajectory of crime has continuously shifted in response to cultural, economic, and technological developments. As human society progresses, so too does the nature of criminal activity, adapting to exploit vulnerabilities within emerging systems (Garland, 2001). With the advent of computer technology, crime expanded into new domains, marking the birth of computer crime and subsequently, cybercrime—a global phenomenon that thrives within the digital landscape, affecting individuals, corporations, and governments alike. However, the integration of artificial intelligence (AI) and the rapid advancement of digital technologies have ushered in what many, including Michael McGuire, describe as a new, more complex stage in the evolution of crime: hypercrime. Defined by its systemic, pervasive, and multifaceted impacts, hypercrime represents a shift beyond cybercrime, exploiting advanced technology to challenge traditional legal and ethical boundaries in unprecedented ways.

This paper examines the historical development of crime, tracing its origins from traditional, tangible forms to digital manifestations in computer and cybercrime, before positing "hypercrime" as the next critical stage in this evolution. Leveraging the insights of McGuire's "Hypercrime: The New Geometry of Harm," the discussion situates hypercrime within a broader social and technological context, arguing that this new paradigm necessitates a significant rethinking of law enforcement's approach. The inherent complexity and vast reach of hypercrime—exacerbated by AI's capabilities—pose profound challenges that render conventional policing methods insufficient. Accordingly, this study advocates for a redefined law enforcement framework, one that is proactive, technologically integrated, and adaptive to the sophisticated risks posed by hypercriminals wielding AI as a tool for harm.

The implications of hypercrime are substantial, threatening not only individual security but the stability of entire digital ecosystems. As AI continues to permeate various sectors, it enhances the potential for criminal exploitation, allowing malicious actors to conduct large-scale, automated, and highly effective attacks. Recognizing hypercrime as a distinct category illuminates the need for more than reactive law enforcement; it calls for a comprehensive strategy rooted in predictive analytics, cross-border cooperation, and robust ethical safeguards. This paper underscores the urgency of evolving our understanding and responses to crime in an AI-driven world, presenting hypercrime as a framework for examining these emergent threats and as a mandate for a new era of crime prevention and mitigation.

Portuguese Translation

O crime, nas suas diversas formas, há muito que reflecte a natureza evolutiva da sociedade. Desde actos rudimentares de roubo e violência nas primeiras civilizações até crimes complexos e orquestrados digitalmente, a trajectória do crime tem mudado continuamente em resposta aos desenvolvimentos culturais, económicos e tecnológicos. À medida que a sociedade humana progride, o mesmo acontece com a natureza da actividade criminosa, adaptando-se para explorar as vulnerabilidades dos sistemas emergentes (Garland, 2001). Com o advento da tecnologia informática, o crime expandiu-se para novos domínios, marcando o nascimento do crime informático e, subsequentemente, do crime cibernético – um fenómeno global que prospera no cenário digital, afetando igualmente indivíduos, empresas e governos. No entanto, a integração da inteligência artificial (IA) e o rápido avanço das tecnologias digitais deram início ao que muitos, incluindo Michael McGuire, descrevem como uma fase nova e mais complexa na evolução do crime: o hipercrime. Definido pelos seus impactos sistémicos, generalizados e multifacetados, o hipercrime representa uma mudança para além do cibercrime, explorando tecnologia avançada para desafiar as fronteiras legais e éticas tradicionais de formas sem precedentes.

Este artigo examina o desenvolvimento histórico do crime, traçando as suas origens desde as formas tradicionais e tangíveis até às manifestações digitais no crime informático e no cibercrime, antes de propor o "hipercrime" como a próxima fase crítica nesta evolução. Aproveitando os conhecimentos do livro "Hypercrime: The New Geometry of Harm" de McGuire, a discussão situa o hipercrime num contexto social e tecnológico mais amplo, argumentando que este novo paradigma necessita de uma repensação significativa da abordagem da aplicação da lei. A complexidade inerente e o vasto alcance do hipercrime — exacerbados pelas capacidades da IA — colocam desafios profundos que tornam os métodos de policiamento convencionais insuficientes. Assim, este estudo defende um quadro de aplicação da lei redefinido, que seja proativo, tecnologicamente integrado e adaptável aos riscos sofisticados apresentados por hipercriminosos que utilizam a IA como ferramenta para causar danos.

As implicações do hipercrime são substanciais, ameaçando não só a segurança individual, mas também a estabilidade de ecossistemas digitais inteiros. À medida que a IA continua a permear vários sectores, aumenta o potencial de exploração criminosa, permitindo que intervenientes maliciosos conduzam ataques em grande escala, automatizados e altamente eficazes. Reconhecer o hipercrime como uma categoria distinta ilumina a necessidade de mais do que uma aplicação reativa da lei; exige uma estratégia abrangente enraizada em análises preditivas, cooperação transfronteiriça e salvaguardas éticas robustas. Este documento sublinha a urgência de evoluir a nossa compreensão e respostas ao crime num mundo impulsionado pela IA, apresentando o hipercrime como um quadro para examinar estas ameaças emergentes e como um mandato para uma nova era de prevenção e mitigação do crime.

## 2. The Evolution of Crime: From Traditional to Cyber

### 2.1 Traditional Crime Origins

Crime has been a pervasive element throughout human history, consistently adapting to the structural and cultural contours of society. In its earliest forms, crime was closely tied to economic necessity, political conflict, and social inequities. Historically, crime has been regarded not just as an act against individuals but as a disruption to communal harmony and order. Economic deprivation often drove individuals towards theft, while political tensions and cultural differences frequently ignited acts of violence, invasion, and territorial disputes. These acts, rooted in tangible motivations and localized interactions, reflect the conditions and priorities of the societies in which they occurred.

Traditional crimes were typically straightforward in nature, easily understood by local communities and adjudicated through direct interventions. These crimes were tangible, often resulting in visible harm to individuals or property, making the perpetrators identifiable and their motives comprehensible within the socio-economic context. Community members played a significant role in maintaining order, with crime addressed through collective or formalized justice systems, such as early tribal councils, town assemblies, or the judiciary in more advanced societies. Punishments were publicly administered, serving as deterrents to uphold social cohesion.

Traditional crime was fundamentally localized, meaning that its impact and resolution were confined to a specific geographic area and cultural setting. The sense of community responsibility for justice and retribution fostered a societal bond, where crime and punishment were intrinsically linked to maintaining local stability. For instance, acts of theft and assault—common forms of crime—were usually witnessed or detected within the community, and responses were swift and direct. Justice systems evolved to manage these breaches, developing codes and penalties to address recurring transgressions, thereby reinforcing social order.

Traditional crime includes offenses such as theft, assault, and murder. These crimes, occurring within a defined social or geographical context, were often managed through rudimentary legal frameworks or community-led justice mechanisms. Theft represented a direct threat to personal or communal resources, necessitating protective measures and retribution, while acts of assault or homicide were dealt with as severe transgressions against personal safety and social harmony. As these societies progressed, more structured legal systems emerged, formalizing the process of dealing with such crimes and establishing standards for evidence, accountability, and punishment.

**2.2 Transition to Computer Crime**
The advent of digital technology in the late 20th century marked a profound shift in the nature of crime, introducing new vulnerabilities and opportunities for exploitation. Unlike traditional crimes, which typically involved physical acts against people or property, computer crime emerged as a distinct category that operated in the digital realm, often with no need for physical interaction. This transition began with the proliferation of computers and electronic communication systems, which transformed personal and professional environments, but also introduced novel risks and vulnerabilities. As computers became integral to financial transactions, government operations, and private data storage, malicious actors saw opportunities to exploit these systems for illicit gain.

The early years of computer crime were marked by significant incidents that highlighted both the capabilities and dangers of the new technology. One of the first documented cases occurred in the 1970s with the spread of "phone phreaking," a practice in which individuals manipulated phone systems to make free calls or disrupt services. While primitive by modern standards, these actions represented a foundational understanding of how electronic systems could be tampered with. This era also saw the emergence of early "hacker" communities, who initially engaged in exploratory hacking out of curiosity but soon moved toward more malicious activities.

A significant milestone in the transition to computer crime occurred in the 1980s, with the rise of computer viruses and other forms of malware. The "Morris Worm," released in 1988, was one of the first worms (B., 2024) to gain widespread attention due to its rapid spread and disruption of thousands of computers connected to the early internet. This incident underscored the vulnerabilities inherent in interconnected systems and signalled the potential for widespread digital harm. The 1980s also saw the advent of computer fraud, where criminals used early internet systems and computer databases to commit identity theft, financial fraud, and data theft. As more institutions and businesses digitized their records, electronic vulnerabilities multiplied, inviting new forms of unauthorized access and manipulation.

The 1990s further cemented the digital frontier as a fertile ground for crime. As email became ubiquitous, phishing attacks emerged, in which attackers would deceive individuals into revealing personal information, such as bank account details or social security numbers. This decade also witnessed the birth of the dark web, providing a concealed network where illicit activities could flourish without easy detection. The emergence of online marketplaces for illegal goods, stolen data, and hacking tools posed a unique challenge to law enforcement, as traditional investigatory methods proved insufficient in this anonymous and decentralized environment.

The emergence of computer crime exposed significant limitations within traditional policing. Law enforcement agencies, trained to handle physical evidence and face-to-face interactions, found themselves unprepared for the complexities of digital crime, which was often conducted remotely, anonymously, and across jurisdictional boundaries (Brenner, 2010). Unlike conventional crimes, computer crime left behind traces that were less tangible, requiring specialized technical skills to detect and interpret digital evidence. Traditional law enforcement lacked both the resources and expertise to address this new category of crime, leading to widespread difficulties in prosecution and prevention.

Recognizing the inadequacy of existing methods, law enforcement agencies began establishing specialized units to combat computer crime. These units were tasked with developing expertise in cybersecurity, digital forensics, and electronic surveillance, enabling them to respond to the unique challenges posed by digital offenders. Agencies partnered with private sector technology firms and educational institutions to train officers in the skills required to identify and combat cyber threats. Additionally, the establishment of early computer crime laws, such as the U.S. Computer Fraud and Abuse Act of 1986, reflected an emerging legal recognition of digital crime and its implications. International cooperation also became increasingly vital, as digital crimes transcended national borders, prompting collaborative efforts through organizations like INTERPOL and Europol.

### 2.3 Cyber Era

With the rapid expansion of internet technology in the late 20th and early 21st centuries, cybercrime emerged as a formalized category of criminal activity that fundamentally redefined traditional notions of crime and security. As individuals, businesses, and governments increasingly embraced the internet for communication, commerce, and data storage, malicious actors found opportunities to exploit these networks for personal gain, ideological motives, or pure disruption. Unlike earlier forms of computer crime, which often focused on isolated incidents involving individual systems, cybercrime evolved as a pervasive, network-based threat that could affect millions of individuals and organizations simultaneously. This era signified a shift toward crimes that were not only digital but also inherently global in nature, challenging the capacity of traditional legal frameworks and law enforcement practices (Wall, 2024).

Cybercrime is distinct from previous forms of crime due to its capacity for distance, anonymity, and scalability. Traditional crimes typically involve physical proximity and direct interaction between the perpetrator and the victim or the victim's property (Yar, 2013). In contrast, cybercrime allows criminals to operate from any location with internet access, targeting individuals and organizations across the globe with ease. This "distanced" nature of cybercrime enables perpetrators to evade identification and apprehension, as they can mask their locations, identities, and even the origin of their attacks.

Anonymity is another defining characteristic of cybercrime, as criminals can leverage tools such as VPNs, encryption, and the dark web to conceal their identities. This anonymity not only complicates efforts to trace cybercriminals but also fosters an environment where criminal activities can thrive without fear of immediate detection or accountability. Furthermore, the scalability of cybercrime allows for attacks that can reach hundreds, thousands, or even millions of victims simultaneously. Unlike traditional crimes, which typically affect single individuals or localized groups, cybercrime can involve mass-scale data breaches, global phishing campaigns, and widespread malware infections, amplifying its impact on society.

Cybercrime encompasses a broad range of illicit activities, including hacking, data theft, identity fraud, financial fraud, and distributed denial-of-service (DDoS) attacks. Hacking remains one of the most prevalent forms of cybercrime, where attackers exploit vulnerabilities in computer systems to gain unauthorized access to sensitive information. Data theft and identity fraud are also common, with cybercriminals targeting personal and financial information to commit theft, extortion, or sell the data on illicit marketplaces. Ransomware attacks, where attackers lock victims out of their systems and demand payment for restoration, have become particularly disruptive, impacting organizations of all sizes, from local businesses to global corporations and even critical infrastructure.

The rise of cybercrime has presented numerous challenges for law enforcement, as traditional policing methods are often inadequate in a digital, borderless environment (Grabosky, 2007). One of the primary challenges is jurisdictional enforcement. Cybercrimes can be committed from any location with internet access, making it difficult to determine which jurisdiction has authority over a case. A cybercriminal in one country can target victims in multiple other countries, creating complex jurisdictional issues that hinder investigations and complicate prosecutions (Finklea, 2015)

Evidence gathering is another significant hurdle in the fight against cybercrime. Unlike physical crime scenes, where tangible evidence can be collected and preserved, cybercrime investigations rely on digital evidence, which is often ephemeral and easily concealed or erased. Cybercriminals can exploit encryption and anonymization tools to hide their tracks, making it difficult for law enforcement to trace activities back to specific individuals. Additionally, the volume of data involved in cybercrime cases can be overwhelming, requiring advanced technical expertise and significant resources to process and analyse.

International cooperation is crucial yet challenging in the fight against cybercrime. As cybercrime is inherently transnational, effective enforcement often requires coordinated efforts across multiple countries. However, differing legal systems, priorities, and capabilities among nations make it difficult to achieve seamless cooperation. Some countries lack the resources or political will to address cybercrime proactively, while others may harbor cybercriminals for political or financial gain. Moreover, international agreements on cybercrime, such as the Budapest Convention on Cybercrime, are limited in scope and are not universally adopted, further complicating collaborative efforts.

In response to these challenges, law enforcement agencies have developed specialized cybercrime units and enhanced their digital forensics capabilities. These units work to stay abreast of the latest cybercrime tactics and technologies, leveraging tools such as artificial intelligence, data analytics, and cross-border intelligence sharing. However, the complexity and rapid evolution of cybercrime continue to outpace law enforcement efforts, highlighting the need for continuous adaptation, collaboration, and innovation.

## 3. From Cybercrime to Hypercrime: Introducing a New Paradigm

### 3.1 Defining Hypercrime in the Context of Michael McGuire's Work

In his influential book Hypercrime: The New Geometry of Harm, criminologist Michael McGuire introduces the concept of "hypercrime" as a new paradigm in the study of criminal behaviour and harm in the digital age. McGuire argues that traditional understandings of crime, particularly those developed to address physical, localized offenses, are insufficient for comprehending the intricacies and impacts of modern, technology-driven criminal activity. He asserts that hypercrime represents a distinct and advanced stage in the evolution of crime—one that transcends traditional boundaries and fundamentally alters how harm is perpetrated, experienced, and perceived in an increasingly interconnected world (McGuire, 2007). By coining the term "hypercrime," McGuire aims to capture the novel, pervasive, and often abstract nature of criminal activities that are deeply embedded in today's digital ecosystem.

McGuire's conceptualization of hypercrime hinges on the idea that modern technologies, especially those associated with the internet, artificial intelligence, and ubiquitous connectivity, have reshaped the ways in which harm can be inflicted. Unlike conventional crimes, which tend to be tangible and localized, hypercrime operates across multiple layers of society and has the potential to inflict harm on a global scale. In this new geometry of harm, digital crime is not confined to specific individuals or locations; rather, it spreads across networks, blurring the distinctions between private and public spheres, between local and international jurisdictions, and between virtual and physical realities. Hypercrime, therefore, is not just an extension of cybercrime but a broader and more insidious form of harm that permeates the very fabric of digital and social spaces.

One of McGuire's central arguments is that hypercrime fundamentally disrupts traditional notions of harm and accountability by transcending conventional boundaries and categories. In the physical world, harm is typically direct, observable, and confined to specific locations or individuals. Hypercrime, however, disrupts this clarity by manifesting in virtual environments where the delineation between victim and perpetrator, harm and intent, is often ambiguous. A single act of hypercrime—such as a sophisticated AI-driven misinformation campaign—can generate ripples of harm across different sectors, affecting individuals, corporations, and governments without a clear or immediate source. This boundary-transcending nature of hypercrime creates a new geometry of harm, one that challenges the core assumptions of existing legal and regulatory frameworks.

Another critical argument McGuire presents is that hypercrime is deeply intertwined with the digital ecosystem itself. In his view, hypercrime does not merely exploit digital technologies but is embedded within the structure and dynamics of the digital world. It exploits inherent vulnerabilities in social media algorithms, artificial intelligence, and data-driven platforms, turning the very tools that enable modern society into vehicles for widespread harm. For example, AI algorithms designed to predict consumer behaviour can be repurposed by hypercriminals to create manipulative deepfakes or to carry out mass-scale identity theft. This interweaving of crime and technology creates a scenario where the digital ecosystem not only facilitates but also intensifies the scale, speed, and complexity of hypercriminal acts (O'Neil, 2016). As Cathy O'Neil (2016) argues, the use of big data and algorithms has the potential to amplify inequalities and obscure accountability, a characteristic feature of hypercrime's systemic and often opaque manipulation of digital technologies.

McGuire further argues that hypercrime fundamentally blurs the lines between the virtual and physical worlds, creating hybrid threats that are difficult to address using conventional law enforcement methods. While cybercrime often focuses on financial or informational damage within digital spaces, hypercrime encompasses broader social, psychological, and even political harms that extend into the physical world. For instance, AI-driven misinformation can erode trust in democratic institutions, causing societal harm that is both virtual and real. Similarly, ransomware attacks targeting critical infrastructure can disrupt essential services, resulting in tangible consequences for communities. By blurring the boundary between the digital and physical realms, hypercrime challenges conventional legal concepts and demands new frameworks that recognize the hybrid nature of these emerging threats.

McGuire's concept of hypercrime underscores the inadequacy of existing legal and enforcement frameworks to address these multidimensional threats. He calls for a rethinking of traditional policing and legal approaches, suggesting that hypercrime requires an adaptive, integrated response that combines technology, policy, and public awareness. The criminal justice system, McGuire argues, must shift from a purely reactive approach to one that anticipates and mitigates potential harms before they manifest. In this sense, hypercrime is not only a call to redefine what constitutes criminality in the digital age but also a mandate to develop innovative strategies for governance and law enforcement.

**3.2 The Case for Hypercrime as the Evolution of Cybercrime**
As digital technologies continue to evolve, so too does the nature of criminal activities associated with these technologies. Hypercrime, as conceptualized by Michael McGuire, is not merely an extension of cybercrime but a new and advanced form of criminality that transcends traditional digital threats (McGuire, 2007). While cybercrime primarily targets individual systems, networks, or data within the digital sphere, hypercrime represents a systemic, all-encompassing threat that leverages advanced technologies—particularly artificial intelligence (AI), the Internet of Things (IoT), and complex networked infrastructures—to inflict widespread harm across digital and physical domains. This shift marks hypercrime as the next evolution of cybercrime, demanding new frameworks and responses from society, law enforcement, and policymakers.

Cybercrime, in its current form, is often defined by attacks on digital assets, typically for financial gain, data theft, or network disruption. The focus of cybercrime is primarily within the digital sphere, where offenders exploit vulnerabilities in networks, software, and databases to carry out activities like hacking, phishing, identity theft, and financial fraud. While the impacts of cybercrime can be extensive, affecting individuals, organizations, and even governments, they are often contained within digital infrastructures and, although disruptive, typically lack direct influence on physical spaces or systemic societal structures.

Hypercrime, however, moves beyond these limitations, operating within and exploiting the increasingly integrated and complex digital-physical ecosystem. Hypercriminals use AI, IoT, and other advanced technologies not merely to hack or steal information but to manipulate entire systems, shape social behaviours, and influence public perceptions on a large scale. AI algorithms, for instance, can be weaponized to create highly personalized misinformation campaigns, capable of manipulating individuals or destabilizing social trust at a scale and precision that traditional cybercriminal methods cannot match. Meanwhile, IoT devices—ranging from smart home devices to industrial control systems—offer hypercriminals access to critical infrastructures, allowing for hybrid attacks that can disrupt both digital networks and physical operations.

In essence, while cybercrime operates within the constraints of the digital environment, hypercrime leverages the interconnectedness of modern society, exploiting weaknesses in both digital and physical networks to create multidimensional threats. These threats do not solely target individuals or isolated systems but can impact broader societal structures, including economic markets, political institutions, and social cohesion. Hypercrime thus represents a paradigm shift from digital offenses to complex, system-wide vulnerabilities that traditional cybercrime does not encompass, marking a profound evolution in the landscape of digital crime.

Implications for Society: The implications of hypercrime are far-reaching, challenging both regulatory frameworks and moral boundaries in unprecedented ways. Hypercrime presents multifaceted risks that threaten social stability, personal security, and economic integrity. One of the most pressing issues posed by hypercrime is its ability to undermine social stability by eroding public trust in institutions and societal structures. AI-driven misinformation campaigns, for example, can destabilize democratic processes by spreading disinformation, influencing voter behaviour, and polarizing public opinion. These campaigns, often indistinguishable from legitimate content, have the potential to manipulate public perceptions, distort reality, and weaken the foundations of democratic governance.

In terms of personal security, hypercrime exploits the expanding network of IoT devices that permeate homes, workplaces, and public spaces. The widespread adoption of IoT has created an unprecedented level of connectivity, but it has also introduced new vulnerabilities that hypercriminals can exploit. For instance, a coordinated attack on smart home devices could compromise individual privacy and security, while a large-scale breach of medical IoT devices could have fatal consequences for patients relying on critical health monitoring. Unlike cybercrime, which is often confined to digital assets, hypercrime directly threatens the physical safety of individuals by leveraging the interconnected nature of modern devices and infrastructures.

Economically, hypercrime presents risks that extend beyond the financial losses typically associated with cybercrime. Hypercrime's systemic impact means that it has the potential to disrupt entire industries, manipulate stock markets, and destabilize economic ecosystems (Berrong, 2018). AI-powered trading algorithms, for instance, could be tampered with to artificially inflate or deflate stock prices, creating financial instability that affects millions of investors. Similarly, attacks on critical infrastructure, such as energy grids or supply chains, could have catastrophic economic repercussions, disrupting the flow of goods and services and impacting the global economy. These actions transcend traditional cybercriminal motivations of profit and instead aim to destabilize economic systems on a macro scale.

Furthermore, hypercrime raises profound ethical and regulatory questions, challenging current laws and moral frameworks. Traditional legal systems are built on clear distinctions between right and wrong, physical harm and digital harm, local jurisdiction and national authority. Hypercrime blurs these boundaries, creating scenarios where harm is systemic and diffuse, perpetrated by anonymous actors across multiple jurisdictions. Regulatory bodies are struggling to keep pace with the rapid technological changes that enable hypercrime, often lacking the tools and expertise necessary to prevent or respond to such threats effectively. This regulatory lag allows hypercriminals to operate with relative impunity, exploiting loopholes and outdated frameworks to further their agendas.

## 4. The Role of AI in Shaping Hypercrime

### 4.1 AI as a Catalyst for Hypercrime
Artificial Intelligence (AI) has emerged as a transformative force across multiple sectors, enhancing efficiency, decision-making, and predictive capabilities (McCarthy, Minsky, Rochester, & Shannon, 2006). However, while AI holds immense potential for innovation and societal benefit, it also serves as a powerful catalyst for hypercrime. In the hands of malicious actors, AI becomes both an enabler and amplifier of hypercriminal activities, facilitating attacks that are increasingly sophisticated, scalable, and automated. Unlike traditional cybercrime tactics, AI-driven threats leverage machine learning, predictive algorithms, and automation to execute hypercriminal activities with unprecedented precision and reach. As AI continues to evolve, its dual-use nature—the capacity to be deployed for both constructive and destructive purposes—complicates efforts to secure digital environments and protect against hypercriminal exploitation (Dehghantanha, 2019)

The application of AI in hypercrime manifests in various forms, each of which leverages the technology's capabilities to bypass conventional cybersecurity measures and exploit human vulnerabilities on a mass scale (Velasco, 2023). One prominent example is **AI-powered phishing schemes**, which use machine learning algorithms to analyse vast amounts of personal data and create highly targeted, contextually relevant phishing messages. Unlike traditional phishing, which relies on generic messaging, AI-enhanced phishing can mimic personalized communication, making it significantly harder for individuals to detect as fraudulent. These schemes can automatically adapt based on user behaviour, refining their techniques to increase the likelihood of success.

Another example is the use of **automated deepfakes** to deceive individuals, manipulate public opinion, or impersonate high-profile figures for malicious purposes. Deepfake technology, powered by AI, can generate realistic video and audio content that is nearly indistinguishable from genuine recordings. Hypercriminals can deploy deepfakes for extortion, blackmail, and misinformation, leveraging the medium's credibility to manipulate perceptions and actions. For instance, AI-generated deepfake videos of political leaders or corporate executives making false statements can have widespread social and economic impacts, eroding public trust and destabilizing institutions.

**Predictive hacking** represents another dimension of AI-driven hypercrime, where AI systems analyse patterns and predict vulnerabilities in security systems. Machine learning algorithms can autonomously scan for weaknesses, learn from previous hacking attempts, and anticipate potential points of entry, significantly enhancing the efficiency and success rate of cyber attacks. Predictive hacking can also be used to design adaptive attacks that evolve in real time, adjusting tactics based on the target's defences. This dynamic, learning-based approach allows hypercriminals to stay ahead of traditional cybersecurity measures, which are often reactive rather than proactive.

Lastly, **self-learning malware** exemplifies the ways in which AI can automate and amplify malicious activities. Unlike conventional malware, which requires human intervention to adapt or update, self-learning malware uses AI algorithms to modify its behaviour autonomously based on the environment it encounters. This adaptability allows the malware to avoid detection by evolving its code to evade antivirus programs, firewalls, and other security protocols. Self-learning malware can be particularly devastating in highly secure environments, as it can continuously modify its structure and behaviour, rendering traditional security defences ineffective.

AI-driven hypercriminal threats introduce new levels of complexity and scalability that challenge existing cybersecurity frameworks. Traditional cyber defences are often rule-based and rely on pre-defined patterns or known signatures to identify malicious activities. However, AI-powered hypercrime operates with a degree of flexibility and unpredictability that conventional defences struggle to counter. Machine learning enables hypercriminal tools to learn from and adapt to security measures, essentially becoming more effective over time. This capability to continuously evolve, coupled with the ability to target multiple entities simultaneously, makes AI-driven hypercrime both resilient and scalable.

For instance, an AI-based phishing campaign can automatically customize messages to millions of individuals, adjusting its strategy based on the responses it receives. Similarly, predictive hacking and self-learning malware can target diverse systems across sectors, with the capacity to learn from each encounter, optimizing future attacks. This scalability allows hypercriminals to achieve a greater impact with fewer resources, amplifying the potential harm. The convergence of complexity and scale in AI-driven threats not only increases the frequency of attacks but also intensifies their potential damage, extending beyond isolated incidents to affect entire infrastructures and societies.

Furthermore, the ability of AI to mimic human behaviour complicates the process of detection and defence. Traditional cybersecurity methods are based on differentiating between human actions and machine-generated anomalies; however, AI-driven hypercrime blurs these distinctions. For instance, deepfake technologies can create digital entities that resemble real individuals, deceiving both humans and automated systems. In financial systems, for example, AI-generated voices or images can be used to bypass security measures that rely on biometric data or voice recognition, leading to severe breaches.

The capacity for hypercriminal AI to outpace existing defences has prompted the need for a paradigm shift in cybersecurity, where reactive approaches are no longer sufficient. Traditional methods such as firewalls, intrusion detection systems, and antivirus software are often inadequate against adaptive, AI-enhanced attacks that do not follow predictable patterns. As AI-driven hypercrime continues to evolve, cybersecurity strategies must likewise evolve, incorporating advanced machine learning, behavioural analytics, and autonomous defence systems capable of responding to hypercriminal tactics in real time.

**4.2 The Weaponization of AI by Malicious Actors**
As artificial intelligence (AI) continues to advance, its dual-use nature—the capacity to serve both constructive and destructive purposes—poses significant ethical and regulatory challenges. While AI has the potential to drive innovation and solve complex problems, it can also be weaponized by malicious actors who exploit its capabilities for criminal purposes. The accessibility of AI tools and the rapid pace of technological advancement mean that hypercriminals are increasingly able to leverage AI for sophisticated attacks, often outpacing the ability of regulatory frameworks to address these threats. This weaponization of AI heightens the risks associated with social engineering, enabling hypercriminals to manipulate individuals and organizations with unprecedented psychological precision. Consequently, the growing availability of AI as a tool for harm necessitates an urgent reconsideration of ethical boundaries and regulatory safeguards to prevent its misuse.

The dual-use nature of AI presents complex ethical dilemmas. On one hand, AI can enhance societal well-being by improving healthcare, advancing scientific research, and streamlining public services. On the other hand, these same capabilities can be exploited for malicious purposes, transforming AI into a powerful tool for hypercriminal activities. The ethical challenge lies in balancing AI's potential benefits with its potential for harm. Traditional ethical frameworks, which often assume a clear distinction between beneficial and harmful technologies, are inadequate for addressing the ambiguity inherent in AI. AI tools that enhance user experience or automate decision-making in legitimate contexts can also be repurposed to manipulate, deceive, or harm when used with malicious intent.

Regulatory challenges compound these ethical dilemmas. As AI technology becomes more accessible, individuals and groups outside of traditional tech and research sectors can acquire or develop sophisticated AI capabilities. Open-source AI models, while promoting innovation and knowledge-sharing, can also provide hypercriminals with the building blocks needed to weaponize AI. The democratization of AI tools means that hypercriminals do not require substantial resources or advanced expertise to deploy AI-driven attacks, as they can adapt pre-existing models to suit their needs. Regulatory bodies face the daunting task of creating policies that curb the misuse of AI without stifling innovation. However, the pace of AI advancement often outstrips regulatory adaptation, leaving significant gaps in oversight. Legal frameworks struggle to define accountability when AI is used for malicious purposes, especially as AI algorithms can operate autonomously and without direct human control, complicating questions of intent and responsibility.

Furthermore, regulatory bodies face the challenge of distinguishing between legitimate AI applications and those with malicious potential. Many AI tools are designed for dual purposes—such as image recognition software or natural language processing—that can be used both for benign applications and for creating deepfakes or automating phishing attacks. This ambiguity complicates enforcement, as blanket restrictions on AI tools could hinder beneficial uses while failing to effectively prevent their misuse by hypercriminals. A comprehensive regulatory approach would require not only technical controls on AI development and deployment but also ethical guidelines and international cooperation, as hypercriminals often operate across borders, exploiting jurisdictional inconsistencies to evade accountability.

One of the most concerning aspects of AI weaponization by hypercriminals is its capacity to enhance social engineering attacks, which manipulate human psychology to exploit vulnerabilities. Social engineering, traditionally involving tactics such as phishing or impersonation, is significantly more effective when augmented with AI. With AI-enhanced manipulation, hypercriminals can craft personalized and contextually relevant messages that increase the likelihood of deceiving targets, heightening the risks of identity theft, misinformation, and extortion.

AI algorithms can analyse vast amounts of publicly available data from social media, online interactions, and digital footprints to build detailed psychological profiles of individuals or organizations. By understanding a target's behavior, preferences, and vulnerabilities, hypercriminals can tailor their approach, making their social engineering efforts far more convincing. For instance, an AI-driven phishing attack can use machine learning to identify a target's recent online purchases, personal interests, or even their communication style. This information enables hypercriminals to create phishing messages that appear authentic and contextually appropriate, increasing the likelihood that the target will respond or disclose sensitive information.

AI-driven social engineering is also highly scalable, allowing hypercriminals to target large groups with tailored messages. In contrast to traditional social engineering, which required significant time and effort to customize attacks for each victim, AI can automate the personalization process, enabling hypercriminals to reach thousands or even millions of targets simultaneously. This scalability not only amplifies the potential damage but also complicates detection and response efforts, as each attack appears uniquely crafted and may not trigger conventional security alerts.

The weaponization of AI for misinformation and manipulation also raises significant concerns. AI algorithms can create realistic deepfakes—videos or audio recordings that appear to feature real individuals, often high-profile figures or organizational leaders. Hypercriminals can use these deepfakes to spread false information, manipulate public opinion, or coerce individuals through blackmail. For example, a deepfake video of a corporate CEO announcing a fraudulent merger could influence stock prices and disrupt markets, causing substantial financial damage. Similarly, a deepfake of a public official making inflammatory statements could incite social unrest, eroding trust in institutions and destabilizing communities.

Extortion schemes are another area where AI enhances social engineering. Hypercriminals can use deepfake technology to create compromising or explicit content that appears to feature the target, using this material to coerce the victim into paying a ransom or performing specific actions. This form of AI-driven extortion preys on personal vulnerability, as the victim may believe that the fabricated content is genuine and fear the social or professional consequences of its release.

## 5. Redefining Law Enforcement's Role: A Proactive, Technology-integrated Approach

### 5.1 Acknowledging the Need for Paradigm Shift in Policing Strategies

The emergence of hypercrime, as conceptualized by Michael McGuire, underscores the urgent need for a paradigm shift in policing strategies. Hypercrime, with its systemic reach, boundary-defying impact, and exploitation of advanced technologies, requires a rethinking of traditional and even contemporary cyber-focused law enforcement approaches. McGuire's arguments highlight how hypercrime transcends the limitations of conventional crime categories, challenging not only operational policing practices but also the fundamental frameworks within which law enforcement agencies conceptualize crime and harm. Integrating McGuire's hypercrime framework into modern policing necessitates an acknowledgment of these new complexities and the development of adaptive, proactive strategies that can address the unique challenges posed by hypercrime (Brown & Marsden, 2015).

McGuire's hypercrime framework advocates for viewing crime in the digital age not merely as isolated, opportunistic acts but as a multifaceted phenomenon that leverages technology to create diffuse and large-scale harm across societies. According to McGuire, hypercrime redefines the geometry of harm by blurring the lines between digital and physical realms, between local and global impacts, and between direct and indirect damage. Law enforcement, therefore, must adopt a similarly expansive approach to conceptualizing and combating crime. Traditional frameworks, which are often based on linear understandings of victim-perpetrator interactions and direct causation, are inadequate for capturing the systemic nature of hypercrime, where harm can be distributed, anonymized, and diffused across vast digital and physical networks.

Incorporating McGuire's framework into modern law enforcement would mean recognizing hypercrime as a distinct category within criminal typologies, distinct from cybercrime due to its scale, sophistication, and societal impact. Rather than treating hypercrime as an extension of existing cybercrime, law enforcement agencies would benefit from treating it as a phenomenon with unique characteristics that require specialized approaches. This conceptual shift would also entail a more collaborative approach between law enforcement and sectors such as technology, psychology, and social sciences to understand the multi-dimensional nature of hypercriminal threats. By embedding McGuire's insights into training, policy-making, and operational practices, law enforcement agencies can develop a more comprehensive understanding of hypercrime, enhancing their ability to anticipate and mitigate its impact.

Traditional policing models are fundamentally reactive, designed to address localized, identifiable crimes with clear physical evidence and direct connections between offenders and victims. This approach has proven effective for centuries in dealing with conventional crimes, but it falls short in the context of hypercrime, where the dynamics are far more complex. Even cybercrime-focused policing strategies, which were developed to tackle offenses within the digital sphere, struggle to address hypercrime's scale, sophistication, and systemic impact. The inadequacy of traditional and cyber-focused models in addressing hypercrime arises from several key challenges.

Firstly, hypercrime's boundary-defying nature complicates jurisdictional enforcement. Traditional policing models are rooted in geographically defined jurisdictions, where law enforcement agencies have authority over specific territories. Hypercrime, however, operates across national and virtual borders, often targeting multiple countries or regions simultaneously. A single hypercriminal act, such as a coordinated misinformation campaign or a ransomware attack on critical infrastructure, can impact victims across different legal jurisdictions, complicating efforts to investigate, prosecute, and prevent such crimes. Without a coordinated, global response, law enforcement agencies are often hampered by fragmented legal frameworks and limitations in cross-border cooperation, which hypercriminals exploit to evade accountability.

Secondly, the intangible and systemic nature of hypercrime challenges traditional evidence-gathering methods. In conventional policing, evidence is typically tangible, whether physical (such as fingerprints or weaponry) or digital (such as logs of unauthorized access). Hypercrime, however, often leaves behind little discernible evidence due to its reliance on anonymized networks, encryption, and self-modifying algorithms. AI-driven attacks, such as those involving self-learning malware or predictive hacking, adapt and evolve, erasing traces of the hypercriminal's presence and complicating forensic analysis. Moreover, the harm inflicted by hypercrime is often indirect, targeting societal trust or institutional integrity rather than specific individuals or organizations, making it difficult to document and attribute responsibility.

Another significant challenge for traditional policing models is the inability to keep pace with the rapid evolution of hypercriminal tactics, which are often enabled by advanced technologies like AI, machine learning, and the Internet of Things. Hypercriminals leverage these technologies to automate, personalize, and scale their attacks in ways that traditional law enforcement models cannot counteract effectively. Law enforcement agencies, bound by bureaucratic constraints and limited resources, frequently lag behind the cutting-edge tools used by hypercriminals. Cyber-focused units may address some digital threats, but the reactive nature of most policing strategies leaves agencies struggling to adapt to the constantly shifting landscape of hypercrime. Without the capacity for real-time threat detection and proactive intervention, traditional and cyber-focused models are largely ill-equipped to prevent or contain hypercriminal attacks before they escalate.

Additionally, the societal impact of hypercrime extends beyond the scope of traditional law enforcement objectives, which are primarily focused on maintaining order and enforcing laws. Hypercrime's capacity to erode public trust in institutions, destabilize social structures, and disrupt economies requires a response that integrates social, psychological, and economic expertise. The challenges posed by hypercrime call for a multi-disciplinary approach, where law enforcement collaborates with stakeholders across technology sectors, academia, and public policy to address the root causes and long-term impacts of hypercriminal activities. Traditional policing models, with their siloed structures and single-discipline focus, are ill-suited to manage these broader societal effects.

**5.2 Proposed Strategies for Mitigating Hypercrime Threats**
In response to the escalating complexity and scale of hypercrime, law enforcement agencies must adopt a multifaceted approach that leverages advanced technologies, promotes international cooperation, and engages the public. The challenges posed by hypercrime, characterized by boundary-transcending impacts and AI-enhanced tactics, demand proactive and innovative strategies that go beyond traditional policing methods. This section outlines proposed strategies to mitigate hypercrime threats through the adoption of advanced technologies, the strengthening of international collaboration, and the promotion of community engagement and public awareness.

To keep pace with the sophisticated methods employed by hypercriminals, law enforcement agencies must integrate advanced technologies into their investigative and preventive practices. AI-based predictive tools, blockchain for traceability, and advanced data analytics are essential components of a proactive approach to hypercrime detection and response.

1. AI-Based Predictive Tools: Hypercriminals often utilize AI to adapt their methods and evade detection, which necessitates the use of equally sophisticated AI tools by law enforcement. Predictive policing algorithms, powered by machine learning, can analyse patterns in data to forecast potential hypercriminal activities and identify at-risk systems or populations. These tools can scan large volumes of data to detect anomalies, anticipate likely targets, and even predict the evolution of ongoing hypercriminal campaigns. By harnessing predictive AI, law enforcement can shift from a reactive stance to a more proactive, preventative approach, reducing the likelihood of successful hypercriminal attacks (Perry, 2013).

2. Blockchain for Traceability: Blockchain technology, known for its transparency and immutability, offers significant potential for enhancing traceability in digital transactions and data transfers. Hypercriminal activities often rely on anonymized transactions, particularly in cases of ransomware payments and illicit online marketplaces. Implementing blockchain for critical infrastructure and high-stakes financial transactions can help law enforcement track digital assets, identify fraudulent behaviour, and prevent the laundering of funds associated with hypercriminal activities. Blockchain's decentralized ledger system can provide a reliable chain of custody for digital evidence, enhancing accountability and aiding in the prosecution of hypercriminals.

3. Advanced Data Analytics: The volume and variety of data involved in hypercrime investigations require sophisticated analytical capabilities. Advanced data analytics, including big data processing, network analysis, and behavioural analytics, can help law enforcement agencies identify hidden connections within hypercriminal networks, detect emerging trends, and uncover patterns indicative of coordinated attacks. By analysing social media, transaction records, and other digital footprints, agencies can gain insights into hypercriminal behaviours, motives, and methods. This data-driven approach supports more efficient resource allocation and strategic planning, empowering agencies to target the most significant threats more effectively.

Given the borderless nature of hypercrime, international collaboration and information-sharing are vital for effective mitigation. Hypercrime often involves actors from multiple jurisdictions, exploiting inconsistencies in international law to avoid detection and prosecution. Strengthening cross-border cooperation and developing shared databases can help law enforcement agencies coordinate their efforts, improve intelligence-sharing, and respond more swiftly to hypercriminal activities.

1. Cross-Border Cooperation: Hypercrime frequently transcends national boundaries, targeting victims and systems across multiple countries. This calls for greater collaboration between national and international law enforcement bodies, such as INTERPOL, Europol, and regional cybersecurity agencies. Joint task forces, specialized in hypercrime, can facilitate the pooling of resources, technical expertise, and personnel, allowing for a more coordinated global response. Such cooperation can also enable law enforcement agencies to overcome jurisdictional barriers, conduct joint investigations, and streamline the extradition process for hypercriminal suspects.

2. Shared Databases and Intelligence Sharing: Developing shared databases that aggregate information on hypercriminal incidents, suspects, and methodologies is crucial for enabling timely responses. A centralized database, accessible to authorized agencies worldwide, would allow law enforcement to cross-reference data, identify emerging trends, and link disparate cases that may involve the same hypercriminal actors. Intelligence-sharing protocols should be established to ensure that agencies are informed of new tactics, technological vulnerabilities, and successful mitigation strategies. Such a system would require robust data protection standards to maintain the privacy of individuals and prevent misuse, but with appropriate safeguards, shared databases can greatly enhance law enforcement's ability to combat hypercrime.

Hypercrime, with its psychological manipulation and AI-driven social engineering tactics, poses significant risks to individuals who may not be aware of the methods and motives of hypercriminals. Public engagement and education are critical components of a holistic strategy to mitigate hypercrime, as informed individuals and organizations are less susceptible to deception and exploitation. Law enforcement agencies should focus on raising awareness about hypercrime, promoting safer digital practices, and encouraging vigilance against AI-driven manipulation.

1. Public Awareness Campaigns: Law enforcement agencies can launch public awareness campaigns to educate communities about hypercrime risks, such as phishing scams, deepfake manipulation, and ransomware attacks. These campaigns should emphasize the importance of secure online practices, including password management, two-factor authentication, and data privacy. By making the public aware of hypercriminal tactics, law enforcement can empower individuals to recognize and respond to potential threats, reducing the number of victims and minimizing the effectiveness of hypercriminal schemes.

2. Digital Literacy Programs: Digital literacy is essential in equipping the public to navigate the complexities of an increasingly digital world. Law enforcement, in partnership with educational institutions and technology companies, can implement programs that teach digital literacy skills, including identifying misinformation, verifying online sources, and understanding AI manipulation tactics. Such programs would be especially valuable for vulnerable groups, such as the elderly or those with limited technological experience, who may be more susceptible to hypercriminal exploitation.

3. Community Partnerships and Reporting Mechanisms: Law enforcement agencies can establish partnerships with local businesses, schools, and community organizations to spread awareness and build resilience against hypercrime. Encouraging communities to report suspicious activities through dedicated hotlines, apps, or online portals can provide valuable intelligence and help law enforcement respond quickly to emerging threats. Additionally, fostering a culture of vigilance and collective responsibility can strengthen community defences, making hypercriminal activities more difficult to perpetrate undetected.

**5.3 Ethical and Legal Considerations in AI Policing**

As law enforcement agencies adopt AI to combat hypercrime, they must navigate complex ethical and legal considerations. AI policing introduces new challenges, particularly in balancing effective surveillance with the protection of civil liberties. Deploying AI in policing can enhance law enforcement's ability to detect, prevent, and respond to hypercrime, but it also raises questions about privacy, consent, and accountability (Ferguson, 2017) (Joh, 2019) (Mittelstadt, Allo, Taddeo, Wachter, & Floridi, 2016). Addressing these ethical concerns requires careful oversight and a commitment to transparency. Additionally, there is a need for policy reforms that adapt existing legal frameworks to support law enforcement's evolving role in addressing hypercrime, while also safeguarding individual rights and promoting ethical use of AI.

The deployment of AI in policing poses significant ethical questions about privacy, consent, and the potential for overreach (Rademacher, 2019). While AI can enable predictive policing, automated surveillance, and real-time threat detection, these capabilities must be carefully regulated to prevent abuses and protect individual freedoms.

1. Privacy and Consent: AI-driven surveillance systems can monitor vast amounts of data, including personal information from social media, location tracking, and biometric data. This level of surveillance, if unchecked, could infringe on individuals' right to privacy and lead to a society where citizens are constantly monitored. The challenge lies in balancing the need for security with respect for privacy, ensuring that surveillance measures are proportionate and targeted. Law enforcement should employ AI-driven tools selectively, focusing on specific threats rather than engaging in broad, indiscriminate monitoring. In cases where personal data is collected, individuals should be informed and, where possible, their consent obtained, especially in less urgent contexts.

2. Accountability and Oversight: The use of AI in policing also raises questions about accountability. Unlike traditional law enforcement methods, AI algorithms often operate with limited human intervention, making it difficult to attribute responsibility if the AI makes an error or leads to unjust outcomes. For instance, an AI-based predictive policing system might disproportionately target certain communities, reinforcing biases without adequate oversight. To mitigate these risks, law enforcement agencies should establish transparent oversight mechanisms to monitor the use of AI, including independent review boards to evaluate the ethical implications of AI deployment. Regular audits and public reporting on AI usage in policing can help build trust, ensuring that these technologies are used responsibly and fairly.

3. Bias and Fairness: AI algorithms are susceptible to bias, which can lead to unfair treatment of certain groups. If an AI system is trained on biased data, it may disproportionately identify individuals from particular demographics as potential suspects, perpetuating discriminatory practices in law enforcement. To address these concerns, law enforcement agencies should work to minimize bias in AI tools by using diverse and representative data and conducting regular audits to identify and correct any discriminatory patterns. Additionally, transparency in algorithmic decision-making, including providing explanations for AI-generated insights, can help ensure that policing practices remain just and equitable.

To support law enforcement's efforts to combat hypercrime while addressing these ethical considerations, a series of legal reforms are necessary. These reforms should promote AI transparency, safeguard digital identities, and strengthen international collaboration on cybercrime.

1. AI Transparency Requirements: Transparency is essential to maintain public trust in AI-based policing. Policymakers should implement regulations that require law enforcement agencies to disclose the AI technologies they use, as well as their intended purposes, data sources, and potential impacts on privacy and civil liberties. These transparency requirements could include mandating the publication of impact assessments that outline the potential risks and benefits of specific AI tools. Moreover, algorithms used in law enforcement should be subject to regular, independent audits to ensure they meet ethical standards and do not introduce biases that could harm marginalized communities. Such measures would enable greater accountability, ensuring that AI is used in a manner that aligns with public interest and ethical principles.

2. Digital Identity Protections: The rise of hypercrime and AI-driven social engineering emphasizes the need for robust digital identity protections. Lawmakers should establish regulations to protect individuals' digital identities, including stricter standards for data handling, consent for data use, and measures to prevent identity theft. These protections could involve creating secure digital identity frameworks that allow individuals to authenticate their identities in digital spaces without exposing personal information to unnecessary risks. Digital identity protections would also support efforts to combat hypercrime by making it more difficult for hypercriminals to impersonate or exploit individuals' identities online. Strengthening digital identity security could reduce vulnerabilities that hypercriminals often exploit, such as through AI-generated deepfakes or phishing schemes.

3. International Cybercrime Laws and Collaboration: Given the transnational nature of hypercrime, international cooperation is essential for effective law enforcement. Current cybercrime laws vary widely between countries, creating legal gaps that hypercriminals exploit. Policymakers should work toward establishing international agreements on hypercrime, similar to the Budapest Convention on Cybercrime, but updated to address the unique challenges posed by AI-enhanced threats. These agreements could include provisions for data-sharing protocols, mutual legal assistance, and extradition for hypercriminal activities, allowing for more seamless collaboration between countries. A standardized international framework for combating hypercrime would ensure that law enforcement agencies have the legal backing to pursue hypercriminals across borders, closing jurisdictional loopholes and promoting global security.

4. Rights to Appeal and Redress: AI-based policing decisions, particularly in predictive policing or risk assessment, can have serious consequences for individuals, potentially leading to unwarranted scrutiny or intervention. Policymakers should ensure that individuals have the right to appeal or seek redress if they believe they have been unjustly targeted by AI-driven policing tools. This right to appeal could include access to information about the AI's decision-making process and an opportunity for individuals to correct any inaccuracies in the data used. By establishing clear avenues for redress, policymakers can help prevent potential abuses and reinforce accountability in AI policing.

5. Ethical AI Standards for Law Enforcement: Governments should develop ethical standards for the use of AI in policing, ensuring that these technologies are deployed in a manner consistent with human rights and democratic values. These standards could cover issues such as data privacy, algorithmic bias, and the proportionality of AI interventions. Developing such standards would require consultation with technology experts, ethicists, law enforcement professionals, and community representatives to create guidelines that reflect diverse perspectives and values. Ethical standards for AI policing would provide a framework for responsible AI use, helping law enforcement agencies navigate complex ethical challenges while maintaining public trust.

**Conclusion**

This paper set out to explore the evolution of crime, tracing its journey from traditional forms through the realms of computer crime and cybercrime, culminating in the concept of hypercrime—a newly emerging frontier characterized by its systemic and multifaceted nature. As elaborated through Michael McGuire's concept of "Hypercrime: The New Geometry of Harm," this study reveals the transformative nature of crime as it increasingly leverages advanced technologies such as artificial intelligence (AI) to amplify harm across both digital and physical realms.

The key findings indicate that hypercrime presents distinct challenges due to its exploitation of technological interconnectedness, use of AI, and its ability to transcend traditional boundaries between digital and physical spaces. Unlike earlier iterations of crime, hypercrime affects entire systems rather than isolated targets, necessitating a comprehensive rethinking of how law enforcement approaches crime prevention and response. Hypercrime, by employing AI-driven misinformation, predictive hacking, and deepfakes, has redefined not only the methods but also the scale and impact of criminal activity.

These findings underscore the need for an evolved law enforcement paradigm—one that integrates predictive analytics, international collaboration, and ethical oversight into everyday policing. The broader significance of this research lies in recognizing hypercrime as a distinct evolution of cybercrime, which requires a shift from traditional reactive strategies to proactive, adaptive methods that are capable of pre-emptively identifying and mitigating AI-driven threats. Addressing hypercrime demands not only technological upgrades but also a holistic strategy that combines technical expertise, legal reforms, and community engagement.

However, the limitations of this study include its reliance on existing literature and the absence of longitudinal data, which could further reveal hypercrime's long-term impacts on societal structures. Additionally, the constrained availability of certain datasets prevented a more granular analysis of specific hypercriminal incidents and their systemic repercussions.

Future research should focus on examining the long-term social consequences of hypercrime, particularly its impact on trust in public institutions and democratic processes. Additionally, the development of standardized global policies for hypercrime mitigation, encompassing digital ethics and AI governance, would be a valuable direction for further study. Such research should also emphasize the collaboration between technology sectors, government bodies, and law enforcement to forge cohesive, international strategies against hypercrime.

Ultimately, addressing hypercrime necessitates innovative approaches that fuse technological advances with ethical safeguards to ensure the security and stability of our increasingly interconnected society. The onus is on law enforcement, policymakers, and technology developers to forge a future where AI's benefits outweigh its risks, and where proactive, ethical oversight serves as a bulwark against emerging threats. Hypercrime presents a challenge that cannot be ignored—it is both a call to action and an opportunity to rethink the very fabric of law enforcement in the digital age.

Wall, D. S. (2024). *Cybercrime: The Transformation of Crime in the Information Age.* John Wiley & Sons.

Yar, M. (2013). *Cybercrime and Society.*

## Acknowledgment

The author thanks Rogerio Meirelles for his contributions in the research.

## Contributor Profile

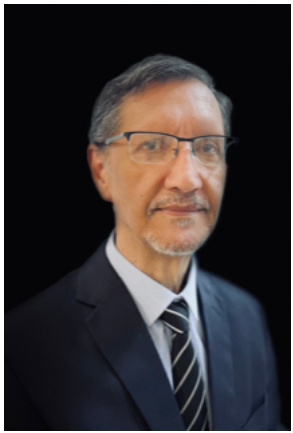

**Rogerio Meirelles** is a retired Federal Agent from the Brazil Federal Police. As a Federal Police Agent Rogerio' public career centered on technological innovation in public security, culminating in his role as Information Technology Coordinator at the Federal Police Department, Brazil. Additionally, he held the post of Deputy Police Attaché at the Brazilian Embassy in Rome, Italy. Following a distinguished 30-year career in policing, he transitioned to private sector work, where he focuses on training and specializing police officers and government employees in Brazil and abroad on advanced investigative tools for digital environments. He also volunteers to give educational lectures to children and teens on responsible social media usage.

## Author Profile

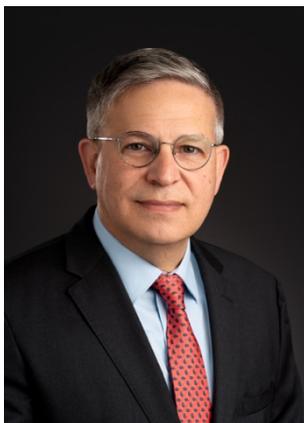

**Francesco (Frank) Schiliro'** is a lecturer in Cybercrime and Cyber Security at ADFA/University of New South Wales, holding a Master of Research in Computing and a PhD in Computer Science from Macquarie University. A retired Superintendent from the Australian Federal Police (AFP), Frank's 34-year policing career began with the New South Wales Police Force in 1988, where he worked in criminal investigations before joining the AFP in 2003.

Frank's research at the University of Macquarie is dedicated to enhancing police effectiveness in crime response through advanced technologies such as artificial intelligence. His current academic focus is on the evolution of crime from cybercrime to hypercrime—an advanced and systemic form of criminal activity that leverages technologies like artificial intelligence to disrupt societal and legal norms. Frank's work explores the complexities of hypercrime, offering insights into its impacts and advocating for new law enforcement strategies to address emerging digital threats.

For more details, please visit Frank's profile at [Macquarie University](#).